\documentclass[aip,apl,reprint,longbibliography,superscriptaddress]{revtex4-1}
\usepackage{mathrsfs}
\usepackage{amsmath,gensymb}
\usepackage{amsfonts}
\usepackage{amssymb}
\usepackage{amsthm}
\usepackage{graphicx}
\usepackage{natbib}
\usepackage{color}
\usepackage{hyperref}
\usepackage{bm}
\usepackage[caption=false]{subfig}
\usepackage{verbatim}

\begin{document}
\title{Magnon-squeezing as a niche of quantum magnonics}

\author{Akashdeep Kamra}
\email{akashdeep.kamra@ntnu.no}
\affiliation{Center for Quantum Spintronics, Department of Physics, Norwegian University of Science and Technology, NO-7491 Trondheim, Norway}
\affiliation{Kavli Institute for Theoretical Physics, University of California Santa Barbara, Santa Barbara, USA}

\author{Wolfgang Belzig}
\affiliation{Department of Physics, University of Konstanz, D-78457 Konstanz, Germany}

\author{Arne Brataas}
\affiliation{Center for Quantum Spintronics, Department of Physics, Norwegian University of Science and Technology, NO-7491 Trondheim, Norway}

\begin{abstract}
The spin excitations of ordered magnets - magnons - mediate transport in magnetic insulators. Their bosonic nature makes them qualitatively distinct from electrons. These features include quantum properties traditionally realized with photons. In this perspective, we present an intuitive discussion of one such phenomenon. Equilibrium magnon-squeezing manifests unique advantageous with magnons as compared to photons, including properties such as entanglement. Building upon the recent progress in the fields of spintronics and quantum optics, we outline challenges and opportunities in this emerging field of quantum magnonics.
\end{abstract}

\maketitle



The spin excitations of ordered magnets, broadly called ``magnons'', carry spin information~\cite{Kruglyak2010,Uchida2010,Adachi2013,Bauer2012,Chumak2015,Cornelissen2015,Goennenwein2015,Lebrun2018} and offer a viable path towards low-dissipation, unconventional computing paradigms. Their bosonic nature enables realizing and exploiting phenomena not admitted by electrons~\cite{Rezende1969,Demokritov2006,Giamarchi2008,Sonin2010,Takei2014,Duine2015,Flebus2016}. The field of ``magnonics'' has made rapid progress towards fundamental physics as well as potential applications in the recent years~\cite{Kruglyak2010,Bauer2012,Chumak2015}. Several studies have also emphasized the quantum nature of magnon quasiparticles resulting in the spin-off entitled ``quantum magnonics''. In this perspective, we outline some recent insights and emerged opportunities focusing on the specific topic of equilibrium magnon-squeezing~\cite{Kamra2016A,Kamra2017A,Kamra2019,Zou2020}. There are many other exciting advancements in the field~\cite{Huebl2013,Zhang2014,Tabuchi2015,Kusminskiy2016,Harder2018} that we will not discuss further here. An overview of these can be found in recent review articles~\cite{Lachance-Quirion2019,Wang2020}. Since the terminology - quantum vs.~classical - sometimes depends on the criterion chosen, we briefly mention the latter as a footnote at appropriate places while employing the term ``quantum'' in our discussion. 

A brief comparison between the fields of quantum~\footnote{The field of quantum optics often uses certain mathematical conditions to characterize a specific state, described by certain wavefunction, of photons as quantum or ``nonclassical''~\cite{Gerry2004,Walls2008}. For example, if the Glauber-Sudarshan P function describing a given state is negative or more singular than a delta function anywhere in the phase space, such a state is ``nonclassical''.} optics~\cite{Walls2008,Gerry2004} and magnonics is in order, since the ideas to be discussed here take inspiration from the former field. While photons and magnons are bosonic excitations described by similar theoretical toolboxes, crucial differences in their physical properties make them complementary in terms of experimental platforms and parameter regimes. We limit the discussion here to only two of these distinctions. First, long optical wavelengths make photons suitable for large systems while magnons fit in on-chip nanodevices. Second, photons need external matter to mediate interactions between them, while magnons are intrinsically interacting. A corollary is that photons have much longer coherence lengths, while magnons provide a compact platform for quantum~\footnote{Here, and in the rest of the article, the term ``quantum effects'' has been used to label consequences, such as entanglement, of having a wavefunction constituted by superposition of various states, as will be discussed below. Furthermore, the Glauber-Sudarshan P function for squeezed states discussed here necessarily takes nonpositive values making such states nonclassical and manifest various quantum effects.} effects and manipulation via interactions. This latter point partly allows the unique niche of magnon-squeezing to be discussed here. In essence, the two fields are complementary and can gain from each other.  

We find it convenient to introduce the equilibrium magnon-squeezing physics first and later place it in the context of the more mature and widely known nonequilibrium squeezing phenomenon~\cite{Walls1983,Walls2008,Gerry2004,Schnabel2017}. Magnons in a ferromagnet admit single-mode squeezing mediated by the relatively weak spin-nonconserving interactions~\cite{Kamra2016A,Kamra2017A}, thereby providing an apt start of the discussion. Antiferromagnetic modes manifest large two-mode squeezing, mediated by the strong exchange interaction~\cite{Kamra2017A,Kamra2019}, and are discussed next. Together, these understandings open avenues towards exploiting quantum phenomena in ``classically ordered'' magnets.



\begin{figure}[t]
	\begin{center}
		\includegraphics[width=86mm]{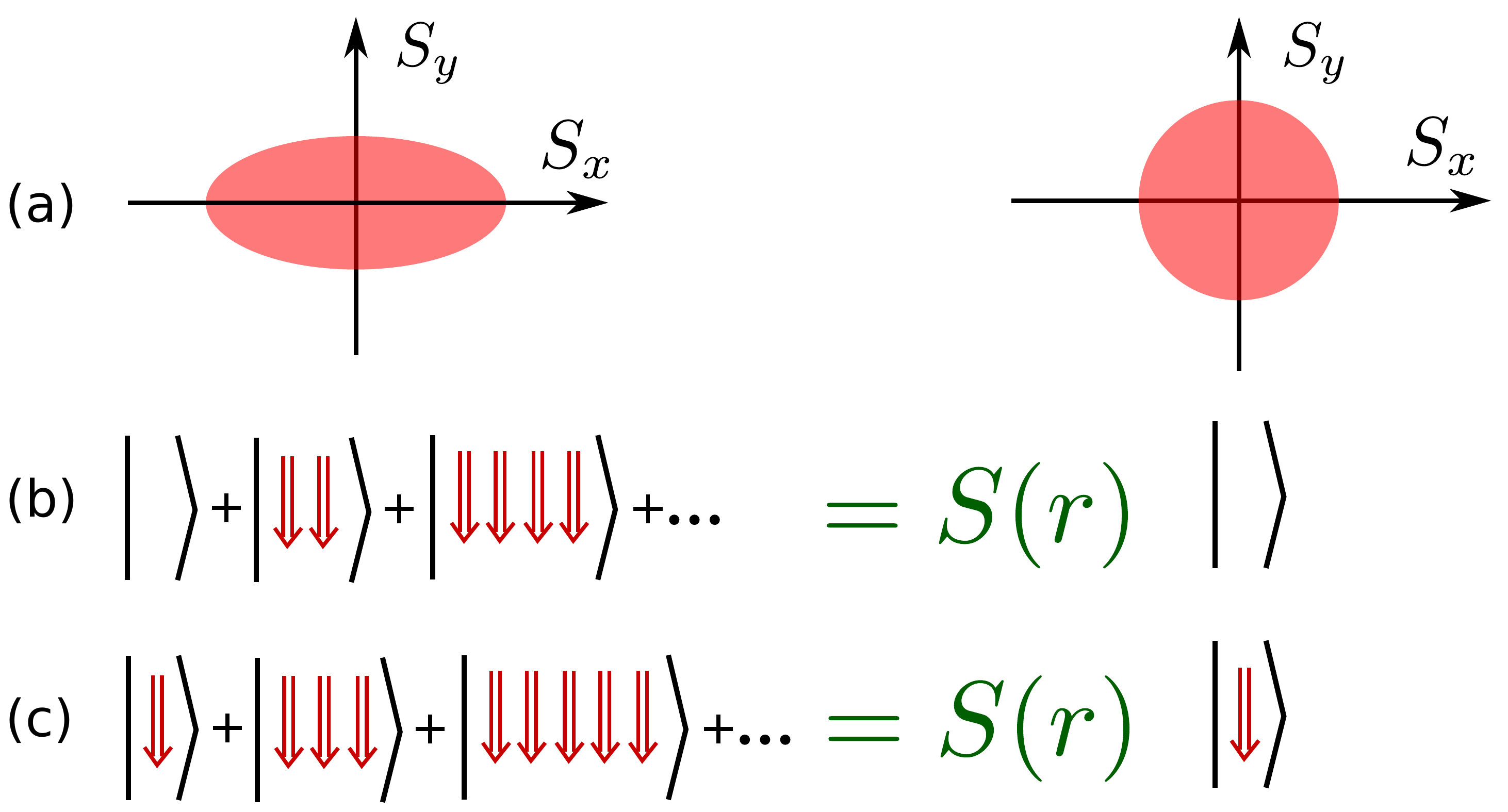}
		\caption{Schematic depiction of the spatially uniform ferromagnetic vacuum and magnon mode in the presence (left) and absence (right) of anisotropy in the transverse (x-y) plane. The ensuing relation between squeezed (left) and unsqueezed (right) modes is also depicted. (a) Heisenberg uncertainty region in the ferromagnetic ground state saturated along the z-axis. The anisotropy in the x-y plane (left panel) causes ellipticity to minimize energy. Schematic depiction of the (b) ground state and (c) magnon-mode wavefunctions for squeezed (left) and unsqueezed (right) ferromagnets related by the squeeze operator $S(r)$. An empty ket and a double arrow respectively denote a fully saturated ferromagnet and a spin-$\hbar$ magnon, which become the ground state and eigenexcitation in the isotropic case (right).}
		\label{fig:ferro}
	\end{center}
\end{figure}

Consider a uniformly ordered ferromagnetic ground state with all the spins pointing along z direction. A spin flip at one of the lattice sites may be seen as a spin $-\hbar$ quasiparticle - magnon - superimposed on the perfectly ordered ground state. In the absence of an anisotropy in the x-y plane, such quasiparticles delocalized in the form of plane waves with wavevectors labeled $\pmb{k}$ constitute the eigen excitations. We focus on the spatially uniform mode corresponding to $\pmb{k} = 0$ that describes the sum over all spins in the ferromagnet. As per Heisenberg uncertainty relation, the total spin in the ferromagnet may not point exactly along the z-direction since that would entail a vanishing uncertainty in both the transverse (x and y) spin components. The latter is not allowed by the Heisenberg principle as the operators for $S_x$ and $S_{y}$ do not commute. Thus, even in the ground state, the total spin manifests quantum~\footnote{The fluctuations being referred to here are a direct manifestation of the Heisenberg uncertainty principle and have no classical analog. They vanish in the limit $\hbar \to 0$ and are understood in terms of a quantum superposition over various configurations (between which the system fluctuates) at the same time. In contrast, a classical fluctuation implies that the system goes through the relevant configurations with time, being in a unique configuration at any given time.} fluctuations and a corresponding uncertainty region schematically depicted in Fig.~\ref{fig:ferro}~(a), right panel. The corresponding wavefunctions for the ground state and eigenmode - magnon - are depicted in Fig.~\ref{fig:ferro}~(b) and (c) on the right.

Now let us include an anisotropy that levies a larger energy cost on the y, as compared to the x, component of the total spin~\footnote{This can be theoretically accomplished by adding a term of the kind $K_y S_y^2 + K_x S_x^2$, with $K_y > K_x  > 0$, to the Hamiltonian. Mathematically, this new contribution necessitates an additional Bogoliubov transformation for diagonalizing the Hamiltonian resulting in a different ground state and eigenexcitations compared to the isotropic case.}. As a result, the system adapts its ground-state quantum fluctuations into an ellipse (Fig.~\ref{fig:ferro} (a), left panel). In this way, it minimizes its energy while obeying the Heisenberg uncertainty principle, which only constraints the area of the uncertainty region and not the shape. The corresponding ground state wavefunction (Fig.~\ref{fig:ferro} (b), left) is constituted by a superposition of the even magnon-number Fock states~\cite{Walls2008,Gerry2004} and is related to the magnon vacuum via the so-called single mode squeeze operator $S(r)$~\cite{Kamra2016A}. Here, $r$ is the so-called squeeze parameter determined for the case at hand by the transverse (x-y) anisotropy. Naively, one can expect the corresponding excitation (Fig.~\ref{fig:ferro} (c), left) to be obtained by superimposing an additional spin flip (magnon) on the vacuum~\cite{Nieto1997,Kral1990} [compare Fig.~\ref{fig:ferro} (b) and (c)]. The resulting eigenexcitation is correspondingly related to the magnon wavefunction via the squeeze operator [Fig.~\ref{fig:ferro} (c)] and is therefore termed squeezed-magnon~\cite{Kamra2016A}. 

The uncertainty region ellipticity, depicted in Fig.~\ref{fig:ferro}~(a), represents a phenomenon distinct from the spin precession ellipticity in the Landau-Lifshitz phenomenology. The former pertains to the shape of the quantum fluctuations around the average spin direction, while the latter describes the trajectory of the average spin in a coherent excited state~\cite{Rezende1969,Glauber1963,Sudarshan1963}. While determined by the same anisotropies for the case at hand, their manifestations and dependencies differ. The squeeze parameter $r~(>0)$ captures the degree of squeezing and the concomitant quantum effects, such as superposition and entanglement. For excitations with high frequencies, the relative energy contribution of the anisotropies becomes small resulting in a diminishing $r$. Furthermore, in contrast with the single-mode case above, modes with $\pmb{k} \neq 0$ manifest two-mode squeezing~\cite{Kamra2016A} that will be discussed in the context of antiferromagnets below. 

To sum up, anisotropies in the transverse plane modify the quantum fluctuations in a ferromagnet [Fig.~\ref{fig:ferro} (a)]. This results in squeezed-vacuum and squeezed-magnon as the corresponding ground state [Fig.~\ref{fig:ferro} (b)] and eigenexcitation [Fig.~\ref{fig:ferro} (c)]~\cite{Kamra2016A}. The anisotropies arise from dipolar or spin-orbit interaction, thereby mediating an effective coupling between magnons required for squeezing~\cite{Kamra2017A}. The effect and importance of anisotropies diminishes with increasing eigenmode frequency and can become relatively weak. Specifically, the noninteger average spin $\hbar^*$ of the squeezed-magnon~\cite{Kamra2016A,Kamra2017A} corresponding to the Kittel mode is $ \gtrsim \hbar$ since the x-y plane anisotropy contribution is typically important. This increase in spin arises from the quantum~\footnote{This is in contrast with a classical superposition, which may be obtained by linearly adding two eigenmodes. For example, a circularly polarized wave may be seen as a classical superposition of two linearly polarized waves. Such a superposition does not mix the different Fock states of the participating modes.} superposition of odd magnon-number states that describes the excitation as depicted in Fig.~\ref{fig:ferro} (c). The average spin approaches $\hbar$ as the eigen excitation frequency significantly exceeds the anisotropy contribution~\cite{Kamra2016B}.



\begin{figure}[t]
	\begin{center}
		\includegraphics[width=86mm]{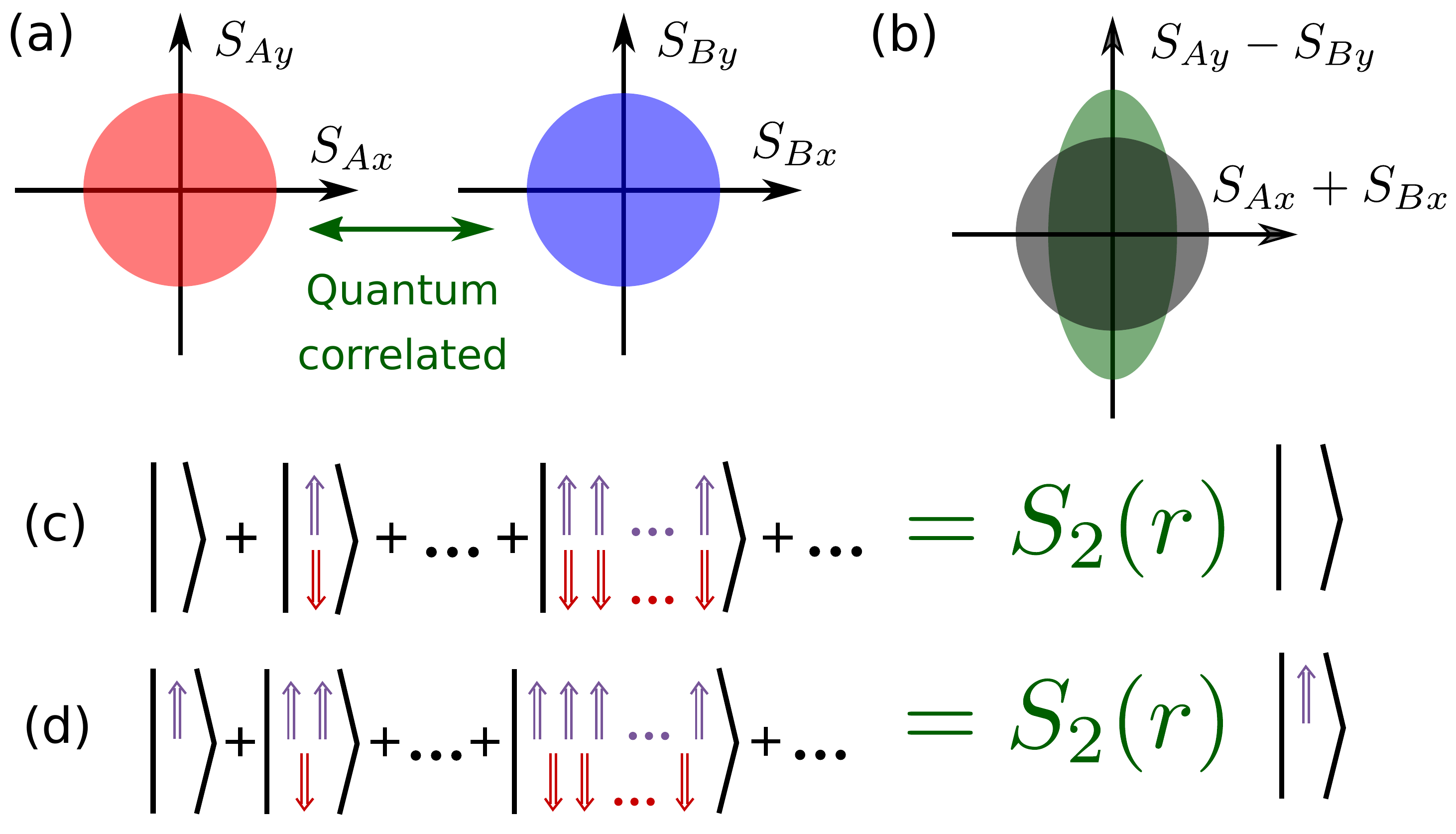}
		\caption{Schematic depiction of spatially uniform antiferromagnetic ground state and eigenmodes. (a) Heisenberg uncertainty regions in the phase spaces of total sublattice spins. The quantum fluctuations in the two sublattices become correlated in order to keep their spins antiparallel to minimize exchange energy cost. (b) The uncertainty region of quantum fluctuations in a combined phase space constructed out of both sublattices. The grey circular region corresponds to the N\'eel state and the green squeezed one represents the actual antiferromagnetic ordered ground state. Schematic depiction of the (c) ground state and (d) spin-up magnon-mode wavefunctions for actual, two-mode squeezed (left) and N\'eel, unsqueezed (right) antiferromagnets related by the two-mode squeeze operator $S_2(r)$. An empty ket denotes the perfectly ordered N\'eel state devoid of any red and blue magnons, that reside on sublattices A and B respectively. }
		\label{fig:antiferro}
	\end{center}
\end{figure}

There are several key differences in antiferromagnets, one being that the strong exchange interaction, and not anisotropy, causes squeezing~\cite{Kamra2019}. Consider a bipartite antiferromagnet in its N\'eel ordered state such that all spins on A (B) sublattice point along (against) z-direction. We disregard anisotropies~\cite{Kamra2017A} and ``turn off'' the antiferromagnetic exchange for the moment. The two sublattices are then equivalent to two isotropic ferromagnets with spins oriented antiparallel to each other. The eigenmodes are spin-down and spin-up magnons residing on sublattice A (red in Fig.~\ref{fig:antiferro}) and B (blue in Fig.~\ref{fig:antiferro}), respectively. As per the Heisenberg uncertainty relation for the total spin ($\pmb{k} = 0$ mode) on each sublattice, the quantum fluctuations in the ground state are now circular in both phase spaces [Fig.~\ref{fig:antiferro} (a)]. The total spins on the two sublattices $\pmb{S}_{A}$ and $\pmb{S}_B$ fluctuate independently, which corresponds to no magnons on either of the sublattices in the ground state.

Now let us ``turn on'' the antiferromagnetic exchange which forces the spins to remain antiparallel. The uncorrelated quantum fluctuations of $\pmb{S}_{A}$ and $\pmb{S}_B$ would cost high energy now as fluctuating independently, $\pmb{S}_{A}$ is not always antiparallel to $\pmb{S}_B$. Thus, mediated by the strong exchange interaction, $\pmb{S}_{A}$ now fluctuates while maintaining its antiparallel direction with respect to $\pmb{S}_{B}$. The system minimizes its energy, while obeying the Heisenberg rule, by bestowing quantum~\footnote{The correlation is quantum in the same sense that the fluctuations are quantum, as discussed above. The correlation is embodied in the states that form the quantum superposition in the total wavefunction. It is also synonymous with entanglement between the two sublattices here.} correlated noise to $\pmb{S}_{A}$ and $\pmb{S}_B$, which individually maintain circular uncertainty regions [Fig.~\ref{fig:antiferro} (a)]. The squeezing now takes place in the phase space constituted by $S_{Ax} + S_{Bx}$ and $S_{Ay} - S_{By}$~\cite{Kamra2019}, as depicted in Fig.~\ref{fig:antiferro} (b). Distinct from the ferromagnet $\pmb{k} = 0$ case, the ground state here is two-mode squeezed where the participating modes are the spin-down and spin-up magnons residing on sublattices A and B~\cite{Kamra2019}, henceforth simply called ``red'' and ``blue'' magnons. The ensuing ground state is formed by a superposition of states with an equal number of red and blue magnons, and is related to the N\'eel state via the two-mode squeeze operator $S_2(r)$ [Fig.~\ref{fig:antiferro} (c)]. The corresponding spin-up eigenexcitation [Fig.~\ref{fig:antiferro} (d)] may be understood as the result of adding a blue magnon to the ground state~\cite{Kamra2019,Kral1990,Nieto1997} [compare Figs.~\ref{fig:antiferro} (c) and (d)].

We considered the $\pmb{k} = 0$ modes above since they admit relatively simple physical pictures. However, the treatment and interpretation for $\pmb{k} \neq 0$ modes are mathematically analogous~\footnote{This holds also for $\pmb{k} \neq 0$ modes in a ferromagnet which manifest two-mode squeezing.} and are implicit in the above two-mode squeezing interpretation differing only in the participating modes and the squeeze parameter $r$, which is wavevector dependent. In antiferromagnets, the squeeze parameter $r$ is large (theoretically divergent for isotropic magnets) for $\pmb{k} = 0$ eigenmodes. It decreases with an increasing k and vanishes as $\pmb{k}$ approaches the Brillouin zone boundary~\cite{Kamra2019}. The squeezing being mediated by the exchange interaction in antiferromagnets bestows them with their unique strong quantum~\footnote{As per our discussion above, the quantum nature of these fluctuations is a direct consequence of the Heisenberg uncertainty relation. However, one may also call these quantum since they cause red and blue magnons to be formed in the ground state at zero temperature thereby diminishing the N\'eel order.} fluctuations and nature.



The notion of squeezing has been developed and exploited in the field of quantum optics~\cite{Walls2008,Gerry2004,Schnabel2017}. For light or photons, the two noncommuting variables, often called quadratures, that embody the Heisenberg uncertainty region are the associated electric and magnetic fields. In this case, the uncertainty region in equilibrium is circular. A squeezing of the fluctuations is achieved by generating pairs of quantum correlated photons via four-wave mixing~\cite{Slusher1985} or parametric down-conversion~\cite{Wu1986}, for example. The ensuing squeezed-state is a transient, nonequilibrium state that decays as the drive is turned off. While such squeezed states are perhaps best known for enabling a beyond-quantum-limit sensitivity of LIGO~\cite{LIGO2011,LIGO2013} that detected gravitational waves~\cite{LIGO2016}, several other quantum properties such as entanglement are inherent to these states and have been studied in great detail~\cite{Ou1992,Ralph1999,Milburn1999,Furrer2012}. 

Similar nonequilibrium squeezed states have also been realized in antiferromagnets~\cite{Zhao2004,Zhao2006,Bossini2019}. By generating correlated pairs of red and blue magnons via Raman scattering with light, experiments observed spin dynamics that could only be explained in terms of a nonequilibrium two-mode squeezed state~\cite{Bossini2019}. The participating modes here are the red and blue magnons with wavevectors at the Brillouin zone boundary, which are the antiferromagnetic eigenmodes on account of a vanishing equilibrium squeezing at these wavevectors. The situation is thus distinct from our discussion of equilibrium squeezing above.



The squeezing perspective and picture presented here capitalizes on insights developed in the field of quantum optics to shed fresh light on magnons, which were investigated~\cite{Holstein1940} four decades before the notion of squeezing was developed~\cite{Walls1983}. This perspective is largely based on the direct mathematical relation between the Bogoliubov transformation~\cite{Holstein1940} and the squeeze operator identified initially in the context of magnon spin current shot noise theory~\cite{Kamra2016A}. The latter could be understood in terms of the noninteger average spin of squeezed-magnons in ferromagnets~\cite{Kamra2016A,Kamra2017A} consistent with the schematic in Fig.~\ref{fig:ferro} (c). This squeezing-based picture of magnons further allows to predict and exploit various effects, such as entanglement~\cite{Kamra2019,Zou2020,Yuan2020,Elyasi2020} and exponentially enhanced coupling~\cite{Kamra2019,Liensberger2019,Erlandsen2019}, already established for light~\cite{Qin2018,Leroux2018,Ou1992,Ralph1999,Milburn1999,Furrer2012} but now in magnetic systems manifesting certain advantages. 

The two key strengths of this magnon-squeezing are (i) its equilibrium nature i.e., squeezing here results from energy minimization, and (ii) large squeeze parameter, relative to what is achieved with light~\cite{Schnabel2017}, due to strong interactions in magnets. These unique features open new avenues. For example, on account of attribute (i), the entanglement inherent to these squeezed states is stabilized against decay by the system's tendency to minimize its energy. Can we design architectures harnessing this entanglement stability for phenomena such as quantum computing and teleportation? This protection from decay is quantified in terms of strong squeezing. How is this stability affected and limited by the dephasing and decoherence processes? The equilibrium nature also makes these phenomena somewhat different from the nonequilibrium squeezing physics encountered in the field of quantum optics. Understanding these differences requires further investigation and will be crucial for exploiting these phenomena towards applications.

The experimental demonstration and exploitation of these effects may capitalize on a large body of knowledge from quantum optics together with a multitude of tools available for solid state systems. Several approaches, such as spin current noise~\cite{Kamra2014,Kamra2016A,Kamra2016B,Matsuo2018,Aftergood2018,Nakata2019,Aftergood2019,Rumyantsev2019,Bender2019}, magnon-photon interaction~\cite{Huebl2013,Zhang2014,Tabuchi2015,Kusminskiy2016,Harder2018,Elyasi2020,Yuan2020,Yuan_arxiv} and NV-center magnetometry~\cite{Taylor2008,Hong2013,Doherty2013,Du2017,Agarwal2017,Flebus2018}, that offer access and control over magnons have emerged in the recent years. These techniques can be broadly classified into (i) detecting the average effect of squeezing-mediated quantum fluctuations, such as the already demonstrated coupling enhancement~\cite{Liensberger2019} or spin dynamics~\cite{Zhao2004,Zhao2006,Bossini2019}, and (ii) those probing the fluctuations themselves. The latter class of methods is expected to offer direct insights into and pathways to exploit the quantum correlations, and constitute an active area witnessing rapid developments.

We conclude this perspective with a geometrical argument underpinning the robustness of phenomena discussed here. Many quantum effects vanish in the limit of $\hbar \to 0$ and the smallness of $\hbar$ underlies their fragile nature. In contrast, the squeezing and related quantum effects result from geometrically deforming the Heisenberg uncertainty region irrespective of its area. Thus, while these squeezing effects fundamentally rely on the Heisenberg uncertainty principle and the corresponding quantum fluctuations, they continue to persist in the limit of $\hbar \to 0$. The robust geometrical nature of equilibrium magnon-squeezing therefore offers unique possibilities towards realizing quantum devices.

We acknowledge financial support from the Research Council of Norway through its Centers of Excellence funding scheme, project 262633, ``QuSpin'', and the DFG through SFB 767. This work was also supported in part by the National Science Foundation under Grant No.~NSF PHY-1748958.

Data sharing is not applicable to this article as no new data were created or analyzed in this study.

\bibliography{QMag}

\end{document}